\documentclass[reprint,
superscriptaddress,
groupedaddress,
 amsmath,amssymb]{revtex4-1}

\usepackage{graphicx}
\usepackage{dcolumn}
\usepackage{bm}
\usepackage[most]{tcolorbox}
\usepackage{hyperref}

\begin{document}


\title{Telecom photon interface of solid-state quantum nodes}

\author{Changhao Li}
\author{Paola Cappellaro}%
\affiliation{ Research Laboratory of Electronics, Massachusetts Institute of Technology, Cambridge, MA 02139}
\affiliation{ Department of Nuclear Science and Engineering,Massachusetts Institute of Technology, Cambridge, MA 02139}

\begin{abstract}
Solid-state spins such as nitrogen-vacancy (NV) center are promising platforms for large-scale quantum networks. Despite the optical interface of NV center system, however, the significant attenuation of its zero-phonon-line photon in optical fiber prevents the network extended to long distances. Therefore a telecom-wavelength photon interface would be essential to reduce the photon loss in transporting quantum information.  Here we propose an efficient scheme for coupling telecom photon to NV center ensembles mediated by rare-earth doped crystal. Specifically, we proposed protocols for high fidelity quantum state transfer and entanglement generation with parameters within reach of current technologies. Such an interface would bring new insights into future implementations of long-range quantum network with NV centers in diamond acting as quantum nodes.
\end{abstract}

\maketitle

\section{Introduction}

Quantum network based on solid-state quantum memories are a  promising platform for long range quantum communication and remote sensing ~\cite{Wehnereaam9288, childress_hanson_2013,RevModPhys.89.035002}. Quantum nodes in a network require robust storage, high fidelity and efficient interface to achieve these demanding applications. Among many physical platforms,  nitrogen vacancy (NV) centers stand out for their very long coherence time in the ground states, making them ideal systems to be used as stationary nodes for quantum computer or sensor networks.  
Microwave and optical interfaces have provided the NV system with a flexible toolset of control knobs, as required in  many emerging technologies.
This has enabled a recent demonstration of   deterministic entanglement generation between two NV centers in diamond~\cite{DeterEntanglement}  with an entanglement rate of about 40 Hz.  
Despite these successes, NV centers present some shortcomings for quantum communication applications: the NV spin has a zero phonon line (ZPL) at 637 nm and it only corresponds to 3 \% of the total emission. 

The propagation loss in optical fibers at this wavelength (8 dB/km) is much larger compared with telecommunication ranges (less than 0.2 dB/km). To extend the entanglement generation scheme to large distance would thus benefit from using  a telecom photon interface.  
The conventional way to overcome this limit is to perform parametric down conversion to convert the photons into telecom-wavelength photons (\textit{tele-photons}), as demonstrated in several systems including quantum dots~\cite{QuantumDot1,QuantumDot2,QuantumDot3},  trapped ions~\cite{TrappedIon,PhysRevLett.120.203601,1901.06317} and atomic ensembles~\cite{AtomicEnsemble2, AtomicEnsemble1,1903.11284}. 

The low emission rate into the ZPL limits the rate of entanglement generation. The emission fraction  into the ZPL could be enhanced by a microcavity via the Purcell effect~\cite{Microcavity2017},  while a  difference frequency generation, used in recent experiments, achieved conversion of single NV photons into telecom wavelength with 17\% efficiency~\cite{PhysRevApplied.9.064031}. However, the signal-to-noise ratio was limited by pump-induced noise in the conversion process~\cite{Ikuta:14,1801.03727} and resonance driving at cryogenic temperature is required, preventing  room temperature applications.

An alternative approach is to work with the microwave interface of NV centers and then up-convert the signal to the desired optical domain. The conversion process can be realized with platforms such as electro-optomechanical~\cite{PhysRevLett.109.130503,NPhyElectOptoMecha,PhysRevLett.114.113601,1901.08228} and electro-optic effects~\cite{PhysRevA.81.063837, PhysRevA.97.052319,1805.04509}, which present strong nonlinearities, but they are usually limited by small bandwidths or low conversion efficiencies.
Atoms or spin ensembles such as rare-earth doped crystals (REDC) are another promising interfaces between optical photons  and microwaves, as they can have both optical and magnetic-dipole transitions.  For example, Er$^{3+}$:Y$_{2}$SiO$_{5}$ possesses an optical transition at 1540 nm which belongs to the C-band telecom range. They can strongly interact with photonic cavities~\cite{Zhong16OE,REDCensemblePhoton17} or microwave resonators~\cite{PhysRevLett.110.157001}.  According to theoretical calculations~\cite{PhysRevLett.113.203601,PhysRevLett.113.063603,1812.03246}, the conversion efficiency could reach near unit under optimal parameters,   and the system has been also explored experimentally~\cite{PhysRevA.92.062313,1712.07735}.  Taking  advantage of the REDC system can be helpful for building NV-based quantum networks with long scales.

In this work,  we propose a scheme for indirect coupling between solid-state qubits (NV center) and telecom photons, with  REDC and microwave (MW) photons serving as intermediate media. The paper is organized as follows: in Sec. \ref{Formalism} we derive the effective interactions of the total system, then in Sec. \ref{Main} we present that with different feasible protocols this hybrid system enables efficient interface between telecom photons and NV centers such as quantum state transfer and entanglement generation. And then we compare the approach here with the photon parametric down-conversion method in Sec. \ref{Discussions}. Finally a short conclusion is given in Sec. \ref{Conclusion}. Our approach would enable more complex operations between telecom photons and NV centers, beyond entanglement generation, in an efficient way and could have interesting applications in future quantum computer or sensor networks based on NV centers in diamond.

\section{Formalism}
\label{Formalism}
The schematic of our proposed hybrid system is depicted in Fig. \ref{Exp_setup}: A REDC is embedded in optical cavity and microwave resonator simultaneously, and a NV center spin ensemble is in the same resonator. The microwave resonator depicted here is a coplanar waveguide at low temperature, and could also be a loop-gap design as described in \cite{PhysRevLett.113.203601,1712.07735}.

\subsection{REDC as a transducer between optical and microwave photons}
\begin{figure}[h]
\centering
\includegraphics[scale=0.42]{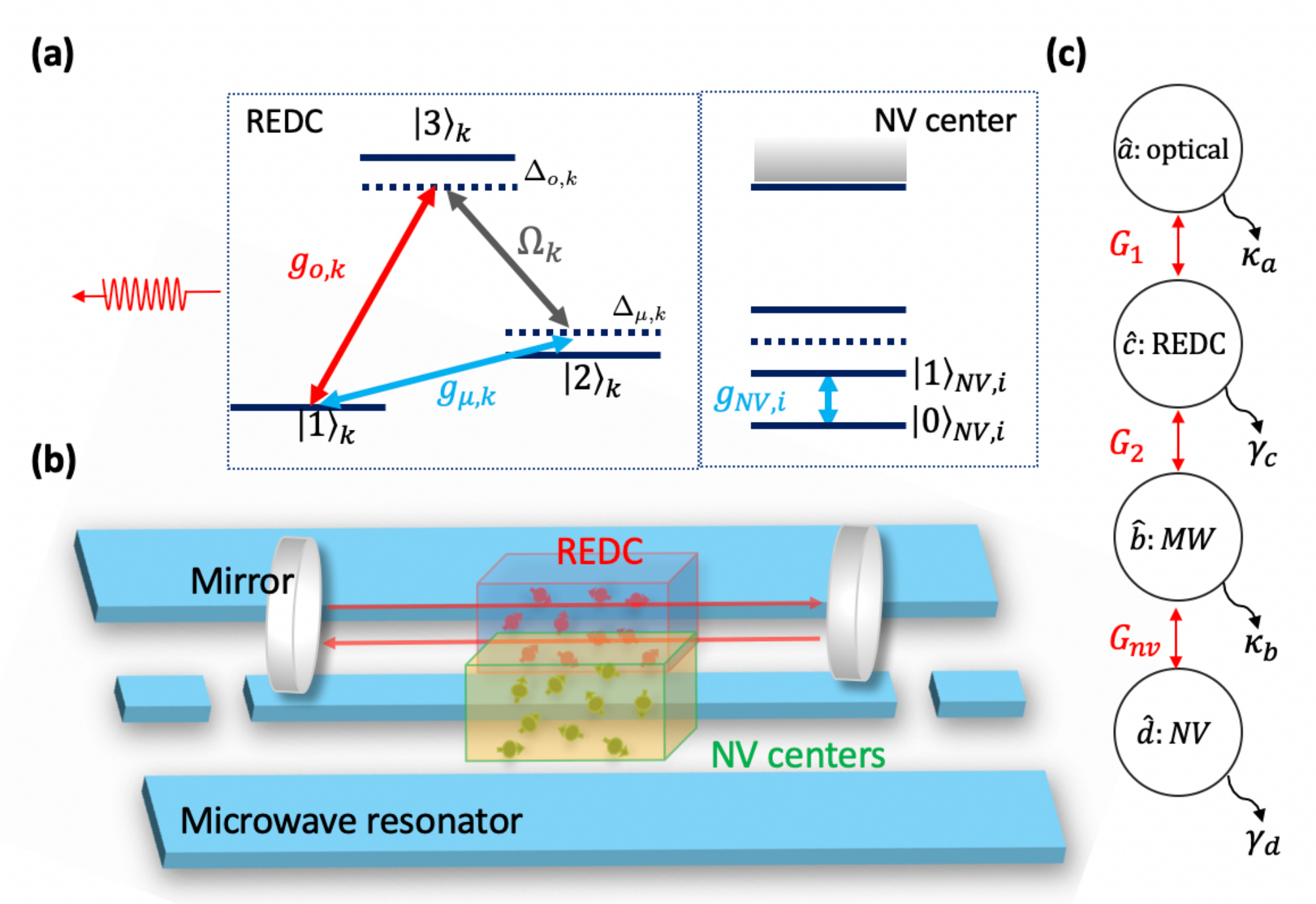}
\caption{\label{Exp_setup}  (a) Effective energy levels of erbium doped crystal and NV centers. The spin k of REDC is driven by an optical field with coupling strength $g_{o,k}$, a microwave field with strength $g_{\mu,k}$ as well as  a classical field with amplitude $\Omega_{k}$. Spin i in NV ensemble is driven by the same microwave field.     (b) Schematic of the hybrid system. A REDC spin ensemble is coupled with optical and microwave fields simultaneously while the NV center ensemble is embedded in the same microwave resonator. (c) Diagrams of the coupled systems and notations for effective coupling strengths and loss parameters.  }
\end{figure}

In Fig. \ref{Exp_setup} (a) we illustrate the energy level of a rare-earth ion. The erbium ion, for example, has an optical transition in the telecom C-band at 1540 nm from $^{4}I_{15/2}$ to the $^{4}I_{13/2}$ (labelled as $|3\rangle$) state.  The $^{4}I_{15/2}$ state further splits into eight Kramers doublets state with transition frequencies in THz range.  Er$^{3+}$ is among the Kramers ions and possesses only a two-fold degenerate spin ground state. At cryogenic temperatures, only the lowest doublet state can be populated and the degeneracy can be lifted by an external magnetic field, resulting in an effective spin 1/2 system. We label the two state as $|0\rangle$ and $|1\rangle$, as denoted in Fig. \ref{Exp_setup}. Compared with other quantum memories, REDC has a large optical depth but also presents large inhomogeneous broadenings  \cite{PhysRevA.91.033834}, thus making usual adiabatic transfer protocols inefficient since it will require the population spends a long time in the spin ensemble. 


We assume $\hbar =1$ and all coupling strengths are real hereafter for simplicity. In the rotating frame of the driven fields, the total Hamiltonian for spins-cavities system is given by:
\begin{eqnarray}
\hat{H}_{\textrm{REDC}}& &= \sum^{N}_{k}(\Delta_{o,k}|3\rangle\! \langle3|_k+\Delta_{\mu,k}|2\rangle\!\langle2|_k)  \nonumber \\
&& +\sum^{N}_{k}(\Omega_{k}|3\rangle\!\langle2|_k + H.c.)  \\
&& +\sum^{N}_{k}(g_{\mu,k}\hat{b}|2\rangle\!\langle1|_k+g_{o,k}\hat{a}|3\rangle\!\langle1|_k+H.c.),\nonumber
\label{totalH}
\end{eqnarray}
where the operators $\hat{a}$ and $\hat{b}$ correspond to the photon mode in the optical and microwave cavity respectively and $N$ is the total number of spins.

Note that the inhomogeneous broadening of transition frequencies can result in random shifts $\delta_{o,k}=\Delta_{o,k}-\Delta_{o}$ and $\delta_{\mu,k}=\Delta_{\mu,k}-\Delta_{\mu}$, where $\Delta_{o}$ and $\Delta_{\mu}$ are average detunings. Here we consider the large detuning regime where $\Delta_{o,k} \gg g_{o,k}, \Omega_{k}, \delta_{o,k}, \delta_{\mu,k}$, and ignore the inhomogeneity of the cavity-spin coupling strength . 
After adiabatic elimination of the excited levels of erbium spins, we obtain the Hamiltonian:
\begin{eqnarray}
\begin{split}
\hat{H}_{\textrm{eff}} =& -\frac{g_{o}^{2}\hat{a}^{\dagger}\hat{a}}{\Delta_{o}}\hat{J}_{11}+(-\frac{|\Omega|^{2}}{\Delta_{o}}+\Delta_{\mu})\hat{J}_{22}  \\
& +((-\frac{g_{o}\hat{a}^{\dagger}\Omega}{\Delta_{o}}+ g_{\mu}\hat{b}^{\dagger}) \hat{J}_{12}+H.c.)   \\
\end{split}
\label{totalH}
\end{eqnarray}

where $\hat{J}_{mn}=\sum_{k}^{N} |m\rangle \langle n|_k$. In the following we will ignore the first two terms as they correspond to nearly homogeneous energy shifts for each spin and could be compensated by tuning the frequencies of the detunings and classical field. The inhomogeneous broadening of optical (microwave) transition for REDC ensemble can be on the order of several  GHz (MHz) \cite{REDCinfo}, for example, 1.5 GHz (3 MHz) as in \cite{PhysRevLett.113.203601}. Working in the large detuning regime makes the REDC robust against inhomogeneity in the optical transition and facilitates controlling the frequency of the output field with broad bandwidth. 


Next, in the low excitation limit, we map the spin ensemble into bosonic modes by introducing the Holstein-Primakoff (HP) approximation \cite{PhysRev.58.1098}:
\begin{eqnarray}
\begin{split}
&\hat{J}_{21}=\hat{c}^{\dagger}\sqrt{N-\hat{c}^{\dagger}\hat{c}}\approx \sqrt{N}\hat{c}^{\dagger}, \\
& \hat{J}_{12}=\hat{c}\sqrt{N-\hat{c}^{\dagger}\hat{c}}\approx \sqrt{N}\hat{c}, \\
& \hat{J}_{z}=\left(\hat{c}^{\dagger}\hat{c}-\frac{N}{2}\right) \\
\end{split}
\end{eqnarray}
where the operators $\hat{c}$ and $\hat{c}^{\dagger}$ obey  bosonic commutation relations approximately. 

With this approximation, we can now obtain the desired, effective Hamiltonian  involving linear coupling between optical mode $\hat{a}$, microwave mode $\hat{b}$ and collective spin wave mode $\hat{c}$:
\begin{eqnarray}
\begin{split}
\hat{H}_{\textrm{eff}} = G_{1}\hat{a}^{\dagger}\hat{c}+G_{2}\hat{b}^{\dagger}\hat{c}+H.c.,
\end{split}
\label{eq:HamEff}
\end{eqnarray}
where $G_{1}=-\frac{g_{o}\Omega \sqrt{N}}{\Delta_{o}}$ and $G_{2}=g_{\mu}\sqrt{N}$ are the collective coupling strengths. 
Thanks to the large optical depths of the ensemble and the cavity-enhanced coupling, the coupling strength $g_{o}\sqrt{N}$ can be on the order of GHz, while $G_{1}$ can be finely tuned by adjusting the detuning $\Delta_{o}$ or the  driving amplitude $\Omega$. 

While we will consider this Hamiltonian as further basis for our analysis, we note that an even simpler description of the microwave-to-optical photon conversion by adiabatically eliminating the mode $\hat c$ that describes the spin ensemble. Indeed, in the limit $\Delta_{\mu,k} \gg |G_{1}|, G_{2},\delta_{\mu,k}$,  one could further adiabatically eliminate the mode $\hat{c}$  and obtain a linear coupling of the form of $\hat{a}^{\dagger}\hat{c}$+H.c.,  with the effective coupling strength further reduced to $|G_{1}|G_{2}/\Delta_{\mu}$. It becomes then clear how the spin ensemble mediates the interaction and we can extract the expected effective rate of the frequency conversion with the input-output formalism \cite{PhysRevLett.113.203601}. To obtain more quantitative prediction of the performance of our scheme, in simulations we retain the more complete Hamiltonian of Eq.~(\ref{eq:HamEff}) and consider experimentally demonstrated parameters.

Strong coupling between collective REDC spin ensemble with microwave cavity can reach 34 MHz  \cite{PhysRevLett.110.157001}  with the inhomogeneous broadening of spin ensemble 12 MHz and resonator decay rate 5.4 MHz. Recent experiment on microwave to optical signal conversion with Raman heterodyne spectroscopy has achieved optical cavity and microwave resonator loss down to less than 10 MHz and 1 MHz respectively \cite{1712.07735}. Hence strong coupling regimes are feasible for current technologies.

\subsection{NV center coupling to microwave photons}
Next we consider the interaction between NV centers and microwave photons. 
We will consider the coupling between the microwave resonator and an ensemble of NV spins, since the coupling to a single NV center is too weak. A spin ensemble is more favorable for the coherent exchange of quantum information:  it has the capacity to store multiple photons and has a simpler experimental realization \cite{PhysRevLett.105.140502,PhysRevA.85.053806}. For example, a coupling strength reaching 16 MHz with resonator decay rate 0.5 MHz (corresponding to Q=3200) and NV ensemble decay rate 10 MHz has been demonstrated \cite{PhysRevA.85.053806} at resonance frequency 2.7 GHz.
In the strong cavity-spin coupling  regime, the decoherence rate of NV ensemble induced by inhomogeneous  broadening can be suppressed due to the cavity protection effect \cite{CavityProtection}. 

For each NV spin $i$, the information can be encoded in two of its triplet ground state levels, for example, $|0\rangle_{i, NV}=|m_{s}=0\rangle$ and $ |1\rangle_{i, NV}=|m_{s}= \pm1\rangle$.  
The NV spin ensemble can be mapped to a bosonic mode by introducing the HP transformation:
\begin{eqnarray}
\begin{split}
&\sum_{i}^{N_{0}}\hat{\sigma}_{NV,i}=\hat{d}\sqrt{N_{0}-\hat{d}^{\dagger}\hat{d}}\approx \sqrt{N_{0}}\hat{d};  \\
&\sum_{i}^{N_{0}}\hat{\sigma}_{NV,i}^{\dagger}=\hat{d}^{\dagger}\sqrt{N_{0}-\hat{d}^{\dagger}\hat{d}}\approx \sqrt{N_{0}}\hat{d}^{\dagger}
\end{split}
\end{eqnarray}
where $\hat{\sigma}_{NV,i}=|0\rangle\langle 1| _{i,NV}$ and $N_{0}$ is the number of NV centers interacting with the microwave cavity. We denote the ground state of the ensemble as $|\{0\} \rangle_{NV}$ which describes all spins in $|m_{s}=0\rangle$. The one-excitation state of the collective wave of the ensemble can be approximated by the symmetric Dicke state $|\{1\} \rangle_{NV}=\sum_{i}^{N_{0}}|00...1_{i}...00\rangle_{NV}/\sqrt{N_{0}}$, where $|00...1_{i}...00\rangle_{NV}$ denotes the state with i-th spin in $|m_{s}= \pm1\rangle$ and the rest in $|m_{s}=0\rangle$.

In the cavity-ensemble resonance case, the interaction between the NV centers and microwave cavity can be directly described by a simple model with an interaction in beam-splitter form, i.e.,
\begin{eqnarray}
\begin{split}
\hat{H}_{NV} = G_{nv}(\hat{b}\hat{d}^{\dagger}+ \hat{b}^{\dagger}\hat{d})
\label{Hnv}
\end{split}
\end{eqnarray}

\subsection{Hybrid System Evolution}
Finally, the Hamiltonian describing the hybrid system can be written in terms of  linear interactions between four bosonic modes:
\begin{eqnarray}
\begin{split}
\hat{H}=\hat{H}_{\textrm{eff}}+\hat{H}_{NV}
\label{HTotal}
\end{split}
\end{eqnarray}
By further considering photon losses and spin decay rates, the hybrid system dynamics is governed by the master equation:
\begin{eqnarray}
\label{master}
\begin{split}
& \frac{d\rho}{dt}=-i[\hat{H},\rho]+\frac{1}{2}\kappa_{a}\zeta(\hat{a})+\frac{1}{2}\gamma_{c}\zeta(\hat{c}) \\
& \frac{1}{2}\kappa_{b}\bar{n}_{th}\zeta(\hat{b})+\frac{1}{2}\kappa_{b}(\bar{n}_{th}+1)\zeta(\hat{b}) +\frac{1}{2}\gamma_{d}\zeta(\hat{d}),
\end{split}
\end{eqnarray}
where $\zeta(o)=2\hat{o}\rho \hat{o}^{\dagger}-\hat{o}^{\dagger}\hat{o}\rho-\rho \hat{o}^{\dagger}\hat{o}$ is the Lindblad operator for a given operator $\hat{o}$ and $\bar{n}_{th}$ is the thermal excitations of the microwave cavity. Here $\kappa_{a(b)}$ is the decay rate of optical (microwave) cavity while  $\gamma_{c(d)}$ is the decoherence rate of REDC (NV center) spin ensemble, as shown in Fig. \ref{Exp_setup} (c) .  Note that the thermal occupations for microwave photons in the frequency range of GHz can be negligible at a temperature around 10 mK. 

It is easier to follow  the system dynamics  by considering the equations of motion of the bosonic operators,
 \begin{equation}
\begin{cases}
\partial_{t}\hat{a}=-iG_{1}\hat{c}-\frac{\kappa_{a}}{2}\hat{a}\\
\partial_{t}\hat{b}=-iG_{2}\hat{c}-iG_{nv}d-\frac{\kappa_{b}}{2}\hat{b}\\
\partial_{t}\hat{c}=-iG_{1}\hat{a}-iG_{2}\hat{b}-\frac{\gamma_{c}}{2}\hat{c}\\
\partial_{t}\hat{d}=-iG_{nv}\hat{b}-\frac{\gamma_{\hat{d}}}{2}\hat{d}\\
\end{cases}
\label{HeisenbergEqu}
\end{equation}
These equations can be further simplified in limiting cases. For example, in the bad cavity limit where $\kappa_{a} \gg |G_{1}|, \kappa_{b}\gg G_{2}, G_{nv}$, the dynamics of spin operators $\hat{c}, \hat{d}$ becomes (Appendix \ref{bad cavity}): 
\begin{eqnarray}
\begin{cases}
\partial_{t}\hat{c}  = -\frac{2G_{2}G_{nv}}{\kappa_{b}}\hat{d} - (\frac{2G_{1}^{2}}{\kappa_{a}}+\frac{2G_{2}^{2}}{\kappa_{b}}+\frac{\gamma_{c}}{2})\hat{c} \\
\partial_{t}\hat{d} =  -\frac{2G_{2}G_{nv}}{\kappa_{b}}\hat{c} - (\frac{2G_{nv}^{2}}{\kappa_{b}}+\frac{\gamma_{d}}{2}) \hat{d}
\end{cases}\label{radiation damping}
\end{eqnarray}
The effective decay rates $\frac{2G_{(i)}^{2}}{\kappa_{a (b)}}$ that appear in addition to the intrinsic decays $\gamma_{c,d}$, describe the radiation damping effect induced by the two cavities \cite{PhysRevA.46.4354}. The overall  decay timescales of the two modes can then be simply evaluated from these equations. This limit could be more easily achieved experimentally, since cavities with low Q factors are only needed, and it would enable fast readout because spontaneous emission of spin ensemble into the cavity mode is effectively an irreversible process. 
However, for the applications in next sections,  we will  focus on the case where coupling strengths are comparable with, or stronger than system dissipations: this enables several excitation exchanges forward and back,  before dissipation has a significant impact.
We will then consider the strong coupling regime which is accessible under current technical capabilities. Note that the ratio of the coupling strengths G$_{1}$/G$_{2}$(G$_{nv}$) can be dynamically controlled and the cavity or spin ensemble resonance frequency can also be  tuned. The system hence  enables  efficient information transfer and operations among the subsystems.

\section{ Quantum state transfer and entanglement generation}
\label{Main}

The form of the effective Hamiltonian we obtained in the previous section, displaying  linear beam-splitter interactions, makes it evident that 
one can use the REDC spin ensemble mode and microwave mode as a bridge to connect the telecom optical photon and the NV spins. In this section, we present protocols for transferring state and generating entanglement between the two subsystems.

\subsection{SWAP protocol for state transfer}
\label{SWAP}

\begin{figure*}
\includegraphics[scale=0.45]{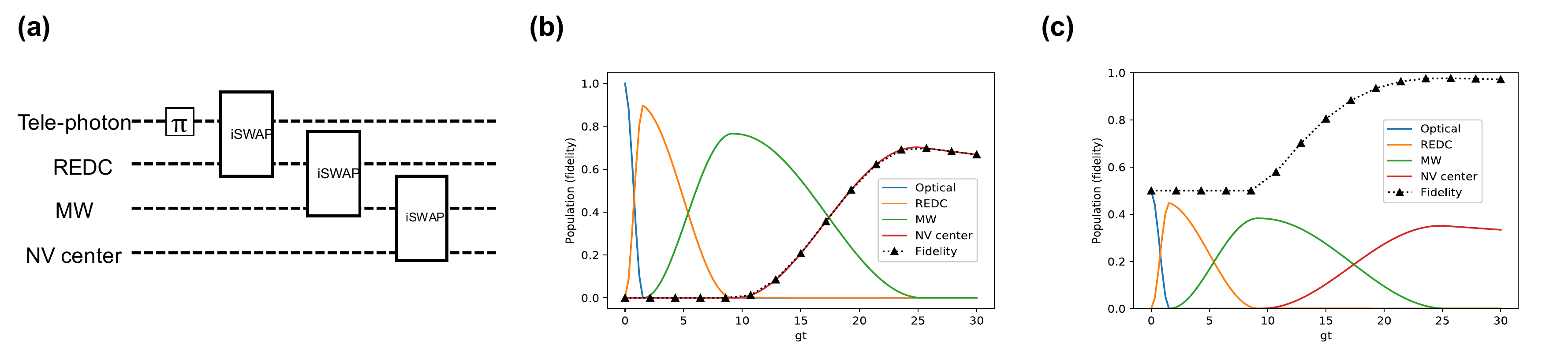}
\caption{\label{Population1} The SWAP protocol for state transfer. (a) Circuits of transferring one photon population to NV centers for the swap protocol.  Initially all systems are in their ground states.   (b)(c) Occupations of subsystems and transfer fidelity as a function of time under system decays . Initially at $t=0$ the optical cavity is in (b) one photon Fock state  and (c) superposition of one and zero photon state, while others are in their ground states. Under subsystem losses, a sequential swap scheme could yield a max fidelity of NV ensemble of over 0.70 in (b) and 0.98 in (c). We take $\kappa_{a}=2.5\gamma_{c}=10\gamma_{nv}=100\kappa_{b}=0.1g$ which are within reach under current technologies as discussed in the main text.}
\end{figure*}

A straightforward state transfer protocol would be based on sequential gates implementing consecutive swaps. 

The first step is to map the photon state to the REDC spin ensemble. Unlike $G_{2}$ and $G_{nv}$, the coupling strength $G_1$ between photons and REDC spins can be dynamically controlled by changing the Rabi frequency $\Omega$ and detuning $\Delta_{o}$.  Note that even in the large tuning limit $\Delta_{o} \gg \Omega, g_{o}$, we can still achieve $|G_{1}| \gg |G_{2}|$, where $G_{2}$ is the collective coupling strength between microwave cavity and spin ensemble, which can be efficiently suppressed when we increase the detuning $\Delta_{\mu}$. At $t=0$, the optical mode is prepared in the state $|\Phi_{o,t=0}\rangle$ (for example, a superposition of Fock states) while all the other systems are prepared in their ground states. Then, for a time  $T_{1}$, we can adjust the effective Hamiltonian of the system to be:
\begin{eqnarray}
\label{Heff1}
\begin{split}
H_{\textrm{eff}} \approx G_{1}(\hat{a}^{\dagger}\hat{c}+\hat{a}\hat{c}^{\dagger})   \quad  0<t< T_{1},
\end{split}
\end{eqnarray}
that is, we suppress all other interactions among the hybrid systems.
The swap time $T_{1}=\frac{\pi}{2|G_{1}|}$ is chosen to correspond to an effective $\pi$ pulse, and can be obtained by simply solving the Heisenberg equations. Note that even if the microwave cavity is far away detuned, the two spin ensembles could still have interactions mediated by virtual microwave photons in the case, for example, that they are in resonance with each other but both detuned from the resonator \cite{PhysRevX.8.041018}, as discussed in Appendix \ref{virtual}.  As long as the detuning is large enough, their effective coupling strength is much smaller than $|G_{1}|$ and this effect can be neglected. 

Alternatively,  without relying on the large detuning limit,  one can use the controlled reversible inhomogeneous broadening protocol (CRIB) to map the photon state to collective atoms \cite{PhysRevLett.113.063603}. However, a spectral hole burning technique is required for this protocol: this not only reduces the effective number of interacting atoms thus decreasing the effective interaction strength, but it also requires a suitable shelving state. Also, long sophisticated pulse sequences are essential to prevent loss of transfer efficiency due to the inhomogeneous broadening induced by gradient magnetic field.

The second step is to transfer the quantum state from the collective REDC spins to the NV spins, with the microwave photons serving as a quantum bus. As described in \cite{PhysRevLett.113.063603}, a standard adiabatic transfer protocol requires that the population spends a long time in the spin ensemble during which the state  would decay due to  inhomogeneous broadening. A straightforward solution is to transfer states step-by-step by controlling the resonator frequencies or switching the external static magnetic fields.   

First, the microwave frequency can be brought into resonance with collective REDC atoms for a time $T_{2} = \frac{\pi}{2G_{2}}$, while the NV spin frequency is far away detuned e.g., by changing the strength of the external magnetic field, and $|G_{1}|$ is tuned to a much smaller value than $G_{2}$. This enables a transfer of quantum state from collective waves in REDC ensemble to microwave photon excitation. 
Then, one can bring the resonator frequency into resonance with NV spin for a time $T_{3}=\frac{\pi}{2G_{nv}}$ while keeping the REDC far detuned to prevent back propagation. This corresponds to another $\pi$ pulse that maps the resonator state to NV spins. Finally, the resonator and NV centers are detuned far away again to avoid disturbance to the NV state. This protocol can be reversed for a state transfer from NV spins to tele-photons. We note that to enable high state fidelities, fast control of the resonator frequency \cite{SCresonatorFreq} and switchable static magnetic field are needed \cite{SwitchBfield}.

We numerically simulate this protocol with the master equation in Eq. (\ref{master}),  including the decay processes of all subsystems. The optical photon is encoded in a superposition of the vacuum state and a single photon wavepacket.  
We take the effective coupling strengths $|G_{1}|=5G_{2}=10G_{nv}=g$ and consider two initial optical photon states $|\Phi_{o,t=0}\rangle=|1\rangle_{o}$, $\frac{1}{\sqrt{2}}$($|1\rangle_{o} +|0\rangle_{o}$). In the one-excitation bases,  the population of the optical photons can be efficiently transferred to the NV ensemble under realistic loss parameters, as shown in Fig. \ref{Population1}. 
To evaluate the transfer performance we calculate the transfer fidelity   $F(t)=(\textrm{Tr}(\sqrt{\sqrt{\rho_{0}}\rho_{NV}(t) \sqrt{\rho_{0}}}))^{2}$, where $\rho_{NV}(t)$ is the reduced density matrix of NV centers at time $t$ (as obtained from the simulation) and $\rho_{0}$ is the ideal  state after the transfer.
The population of each subsystems, as well as the total transfer fidelity as a function of evolution time are shown in Fig. \ref{Population1},  demonstrating the efficiency of this protocol. With the parameters above, we estimate the transfer rate on the order of 1 MHz, which is limited by the coupling strengths.  Note that the parameters we choose in the simulations here and in the following are quite conservative,  and better performance would be expected under optimal experimental conditions.

\subsection{Adiabatic passage state transfer protocol}

\begin{figure}[h]
\centering
\includegraphics[scale=0.32]{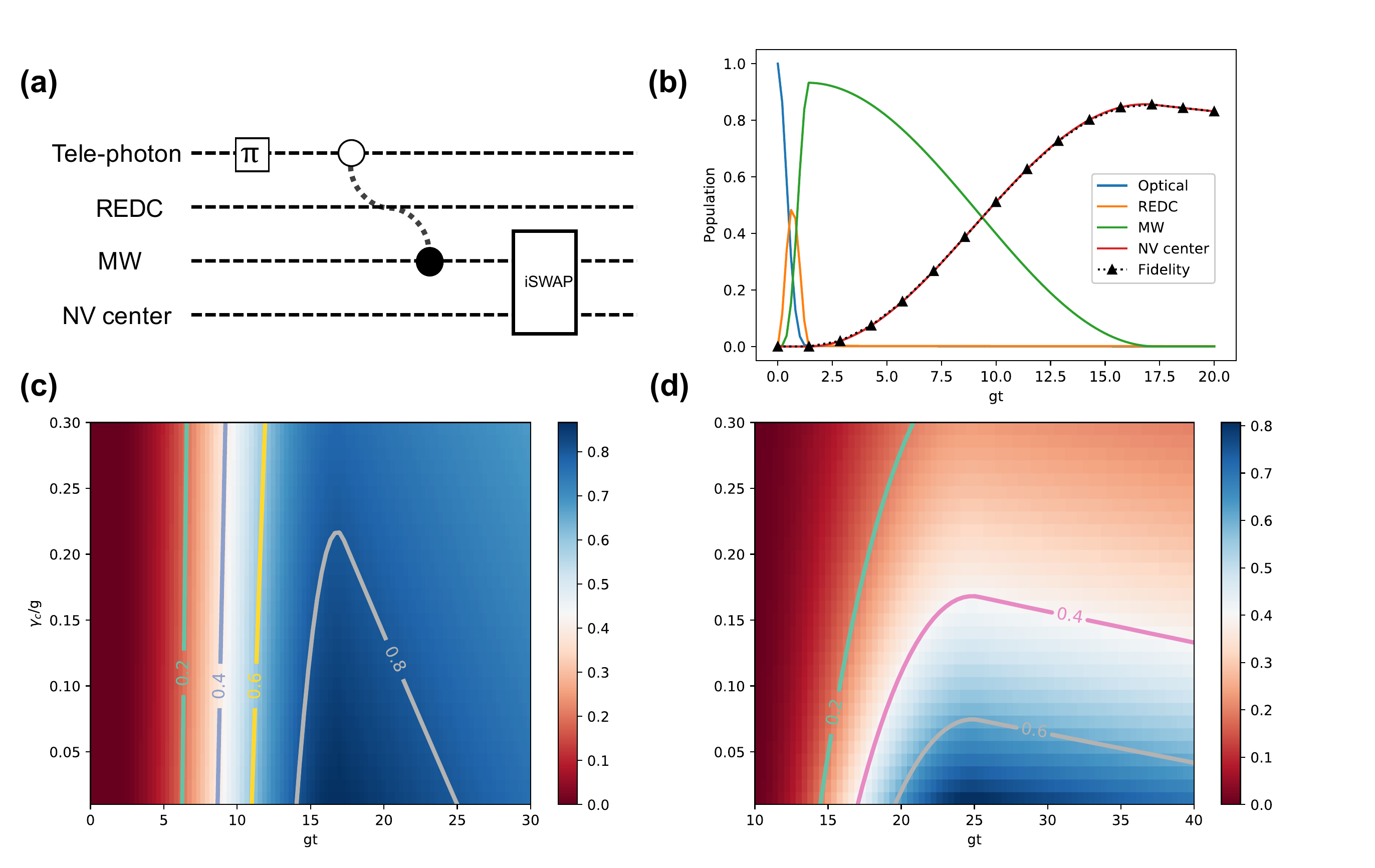}
\caption{\label{DarkState}Adiabatic protocol for state transfer. (a) Circuits of transferring one photon population to NV centers for the adiabatic protocol. An open and a closed circle connected by a dashed line denote an adiabatic state transfer process. (b) Populations of different subsystems as a function of time using the dark-state protocol.  The transfer efficiency is 0.86 here and we take $\gamma_{c}=0.04g$. (c) Transfer fidelity at a function of time for different REDC decay rate $\gamma_{c}$. Initially at $t=0$ the optical cavity is in one photon Fock state while others are in their ground states. As a comparison, in (d) we plot the fidelities in the $\pi$-pulse swap protocol (coupling parameters are the same as in Fig. \ref{Population1}).  In all plots the other decay parameters are the same as in Fig. \ref{Population1}. }
\end{figure}

In the study of REDC-based tele-photon quantum interfaces, a key limiting factor is the large inhomogeneous broadenings of both the optical and microwave transitions \cite{PhysRevLett.113.063603}. While the excited level of spin ensemble has been adiabatically eliminated, the decay loss induced by the microwave transition inhomogeneity can be further eliminated with a dark-state conversion protocol. In addition to the consecutive SWAP method described above, here we suggest an alternative strategy to transfer quantum state between two photonic cavities which is similar to the well-know stimulated Raman adiabatic passage (STIRAP) and has been studied in optomechanics \cite{PhysRevLett.108.153603} and hybrid quantum devices \cite{PhysRevA.96.032342}.

When the microwave resonator frequency $\omega_{b}$ is on resonance with the REDC frequency $\omega_{c}$, whereas the NV center frequency $\omega_{NV}$ is detuned far way due to an applied magnetic field, the total interaction Hamiltonian reads $\hat{H}_{\textrm{tot}}=\hat{H}_{\textrm{eff}}+\sum_{i}\chi_{i} \sigma_{NV,i}^{z}\hat{b}^{\dagger}\hat{b} \approx G_{1}\hat{a}^{\dagger}\hat{c}+G_{2}\hat{b}^{\dagger}\hat{c}+H.c.$ where $\chi_{i}= g_{nv, i}^{2}/|\omega_{b}-\omega_{NV,i}|$ for the i-th NV spin.  Instead of transferring population in a three-level system as in STIRAP, we can introduce system eigenmodes describing quasi-particles formed by hybridization of optical and microwave photons, including hybrid dark, $\hat{h}_{D}$, and bright, $\hat{h}_{B}$, modes:
\begin{eqnarray}
\begin{split}
& \hat{h}_{B}=\sin\theta\, \hat{a} + \cos \theta\, \hat{b}; \\
& \hat{h}_{D}=-\cos \theta\, \hat{a}+ \sin \theta\, \hat{b}; \\
& \hat{h}_{\pm}=\frac{1}{\sqrt{2}}(\hat{h}_{B}\pm \hat{c}),
\end{split}
\end{eqnarray}
where $\tan \theta = |G_{1}|/G_{2} = -g_{o}\Omega / \Delta_{o}g_{\mu}$ can be dynamically tuned by the controlling the detuning $\Delta_{o}$ and the classical field $\Omega$. Now the system can be characterized by the three eigen-modes:
\begin{eqnarray}
\begin{split}
H_{adiab} \approx \omega_{d}\hat{h}_{D}^{\dagger}\hat{h}_{D} + \omega_{+}\hat{h}_{+}^{\dagger}\hat{h}_{+} + \omega_{-}\hat h_{-}^{\dagger}\hat{h}_{-}
\end{split}
\end{eqnarray}
with $\omega_{d}=\omega_{c}$ and $\omega_{\pm}=\omega_{c}\pm \sqrt{\omega_{o}^{2}+\omega_{b}^{2}}$. As one rotates $\theta$ from 0 to $\pm \pi/2$ adiabatically, the dark mode will evolve from $-\hat{a}$ to $\pm \hat{b}$. During this evolution,  the REDC spin ensemble, similar to the intermediate state in STIRAP, remains nearly unpopulated, thus avoiding its decay.
In this step, the state from the cavity mode $\hat{a}$ is transferred to resonator mode $\hat{b}$. Then, the microwave photon state can be transferred to the NV spins by bringing them into resonance as in the SWAP protocol we discussed before. 
Note that the protocol might also be extended to the case where both REDC spin ensemble and microwave photons are adiabatically eliminated and a ``direct'' state transfer between optical photon and NV centers becomes  possible. However, this would require an even longer evolution time thus significantly increasing decoherence effects. 

To evaluate the performance of the adiabatic transfer protocol, we numerically simulate the adiabatic protocol, assuming that initially the cavity is in one photon state while the other subsystems are in their ground states. We consider to  dynamical vary the coupling strength with Gaussian time-dependence, $|G_{1}|=g e^{(t-3)^{2}/15}$  while  the other couplings are fixed, $G_{2}=1.5g=15G_{nv}$. We note that further optimization can be performed in changing $G_{2}$ by tuning the resonator frequency and shaping the time variation of the coupling strengths.
Fig. \ref{DarkState} (b) shows how the REDC will only have a population less than 0.5 during the whole protocol. This leads to the protocol being immune to REDC ensemble losses due to inhomogeneous broadening and a high transfer fidelity can still be generated even when $\gamma_{c}$ is in the order of  or larger than 10 MHz, as shown in Fig. \ref{DarkState}(c). We compare the adiabatic and  $\pi$-pulse SWAP schemes,   demonstrating that the latter will show a significant loss under the same conditions.      Due to this robustness, the adiabatic  scheme  performs best when the REDC ensemble decay is the main loss channel. However, in practice, dissipation in the optical cavity might still induce decoherence thus limiting the fidelity. To speed up the adiabatic process and mitigate the infidelity simultaneously,  shortcuts to adiabaticity \cite{TQD2009, PhysRevLett.116.230503, 1809.00102},  such as transition-less quantum driving, could be applied and would enable fast and robust controls.

\subsection{Entanglement generation between tele-photon and NV spin ensemble}

\begin{figure*}
\includegraphics[scale=0.45]{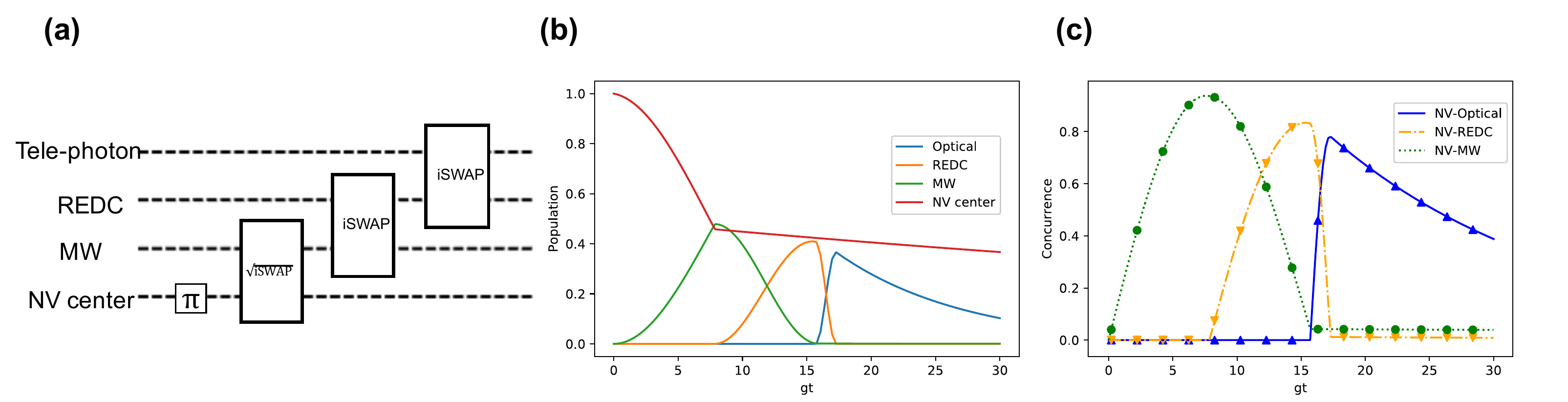}
\caption{ \label{EntangleRev} Entanglement generation between tele-photon and NV spin ensemble. (a) Circuit for generating entanglement. Initially all the systems are in their ground states. (b) Subsystem population as a function of time.  (c) Concurrence between NV centers and other subsystems as a function of time. The final entanglement of NV centers and optical photons can reach a concurrence of 0.78. The coupling strengths are $|G_{1}|=5G_{2}=10G_{nv}=g$ with decay parameters $\kappa_{a}=2.5\gamma_{c}=10\gamma_{nv}=100\kappa_{b}=0.1g$. }
\end{figure*}

To entangle distant  NV centers for applications such as quantum communication or remote sensing, generating entanglement of telecom photons and NV centers will be required. Here, we show that this could be realized in our proposed hybrid system with simple sequential gates.

Consider the Hamiltonian in the one-excitation subspace of the four modes. The evolution of  two neighboring modes $\hat{o}_{i}$ and $\hat{o}_{j}$ (with coupling strength $g_{i,j}$) for a time $t$ will be given by the following matrix:

\begin{eqnarray}
\begin{split}
& U_{i,j}(t)=e^{-i(\hat{o}_{i}^{\dagger}\hat{o}_{j}+H.c)g_{i,j}t}= \\ 
& \begin{pmatrix}
   1 & 0 & 0 & 0 \\
   0 & \cos(g_{i,j}t)  & i\sin(g_{i,j}t) & 0 \\
   0 & i\sin(g_{i,j}t) &  \cos(g_{i,j}t) & 0 \\
   0 & 0 & 0 & 1
    \end{pmatrix}
    \end{split}
\end{eqnarray}

Interaction for a duration  $g_{i,j}t=\pi/2$ will correspond to the iSWAP gate  used in the protocol of Sec.~(\ref{SWAP}), while $g_{i,j}t=\pi/4$ will yield the $\sqrt{iSWAP}$ gate between subsystem $i$ and $j$.

 To create a maximally entangled Bell state one can initially prepare the NV ensemble  in one excitation state while the other systems are in their ground state, and the following protocol can be implemented (Fig. \ref{EntangleRev} (a)): first a $\sqrt{iSWAP}$ gate is applied while switching off $G_{2}$ by detuning the microwave resonator or changing the applied magnetic field, and this would generate entanglement between NV ensemble and microwave resonator; then two consecutive iSWAP gate can transfer the state from the  microwave resonator to the REDC ensemble and then to telecom optical photon.  A step-by-step analysis of the protocol is in Appendix \ref{LossyEntanglementProtocol}. 
Note this protocol could be implemented in reverse, with the excitation initially stored in the optical photon, but this will demand a long photon storage time, and the large decay in optical cavity will significantly worse the final entanglement generation (Appendix \ref{LossyEntanglementProtocol}). 

In Fig. \ref{EntangleRev} we simulate this procedure with the same coupling and decay parameters as in the SWAP state transfer protocol discussed above.  Entanglement between NV centers and telecom photons has a concurrence around 0.8. This could be further improved in practice since the parameters here are quite conservative. 
One can create an arbitrary entangled state $\alpha |1\rangle_{o}|\{0\}\rangle_{NV}+\beta |0\rangle_{o}|\{1\}\rangle_{NV}$ (up to a relative phase factor) by changing the NV-MW interaction time from $\pi/4G_{nv}$ to a $t_{\alpha}$ satisfying $\sin^{2}(G_{nv}t_{\alpha})=|\alpha|^{2}$.

In this protocol the NV-photon entanglement generation rate is on the order of 1 MHz. In principle one could further extend the scheme to create remote NV-NV entanglement by sending the photons to a common station where they are measurement after passing through a balanced  beam splitter \cite{DLCZ}. 
When $\alpha$ is small, the detection of one photon heralds the creation of the maximally-entangled state of two NV ensembles. For a typical photon detection efficiency $p_{det}\sim 10^{-4}$ \cite{DeterEntanglement}, the corresponding generation rate can reach sub-kHZ. As the decoherence rate of entangled NV pairs can be greatly suppressed by control techniques including dynamical decoupling and double quantum driving \cite{PhysRevX.8.031025},  deterministic generation of entanglement might be feasible with our proposed interface \cite{DeterEntanglement}.


\section{Discussions}
\label{Discussions}

The two most important figures of merit of telecom photon-to-matter interface are the rate of entanglement generation (or state transfer) and the fidelity, or signal-to-noise ratio (SNR). In the spontaneous parametric down-conversion (SPDC) approach, applied to NV centers, the expected entanglement rate can be estimated from the probability of detecting a ZPL photon $p_{\textrm{ZPL}}$ per resonant optical excitation and the conversion efficiency.  Recent experiments \cite{PhysRevApplied.9.064031} have  achieved $p_{\textrm{ZPL}} \sim 5.7 \times 10^{-4}$ counts per excitation photon, with a total conversion efficiency of $17\%$. Due to the  broad phonon sideband in the NV emission spectrum, the ratio of ZPL photons sets an upper bound to the achievable rates.

Our protocol, instead, exploiting the microwave interface of NV centers,  is not limited by the ZPL emission rate and does not need  resonant optical driving for NV centers.  The input microwave power and optical pump power could be increased to improve the microwave-optical signal conversion efficiency without the worry of the inverse conversion in the  SPDC method.   Still, as the coupling strengths between subsystems considered here are usually on the order of MHz, the rate of our protocol is  limited by the gate time, which is considerably longer than the optical circle time. In principle, this could be improved in the future by implementing the scheme in ultra-strong coupling regime. 

Another important metric is the final fidelity of the transferred state or the  entanglement fidelity, which can be suppressed by system decoherence and noise.  In the SPDC method, the SNR is limited by both detector dark counts and noise induced by pump laser field. 
As discussed above, our protocol can efficiently perform the tasks under realistic parameters and its conversion efficiency  could reach near unit at low temperature (mK) \cite{PhysRevLett.113.203601,1712.07735}. 
We point out that the protocol fidelities might be reduced by inhomogeneities in the coupling strengths and energy-level detuning,  but these infidelities could be compensated by increasing the Q factors of the cavity and resonator, as discussed in \cite{PhysRevLett.113.203601}. The induced noise would also include detector dark counts and scattering from the classical driving field.  
We note that even if our proposal  involves more subsystems and thus seems to demand a more complex experiment, it does not require spectral filtering to reduce the pump noise as needed in SPDC, while at the same time it could enable more complex gates and operations, besides  entanglement generation. Moreover, even if the microwave resonator should be at low temperature, in principle  the NV centers could be at room temperature, since we don't need resonant driving of optical transitions.

Photon indistinguishability is another important requirement for applications in quantum networks, such as entanglement swapping in quantum repeaters \cite{DeterEntanglement}.  The spectral diffusion of NV centers will result in a frequency difference in the optical transition, and charge fluctuations would further lead to long-term linewidth broadening \cite{SpectralDiff2014}. In our approach, the output photon frequency difference would only depend on the optical cavity which could be improved by increasing the cavity fineness. 


\section{Conclusion}
\label{Conclusion}
We propose a hybrid system to interface an ensemble of NV centers to photons at telecom wavelength. The former can act as a quantum station while the latter can be used as flying qubits for applications in long distance quantum network in fibers without significant loss. We show that with REDC spin ensemble as a medium, we can effectively construct indirect coupling between NV centers and telecom photons. We proposed and numerically test applications to  high fidelity quantum state transfer and efficient entanglement generation. The proposed schemes are within reach of current technologies. Such an interface would open new opportunities into future implementations of long-range quantum network with NV centers in diamond acting as quantum nodes.

\begin{acknowledgments}
We  would like to thank Akira Sone for helpful discussions. This work was in part supported by NSF grant EFRI ACQUIRE 1641064.
\end{acknowledgments}

\appendix

\section{System dynamics in the bad cavity limit}\label{bad cavity}
By integrating the first two equations in Eq. (\ref{HeisenbergEqu}) we get:
\begin{eqnarray}
\begin{cases}
\hat{a}(t) =\int_{-\infty}^{t}e^{-\frac{\kappa_{a}}{2}(t-t^{\prime})}(-iG_{1}\hat{c}(t^{\prime}))dt^{\prime}  \\
\hat{b}(t) =\int_{-\infty}^{t}e^{-\frac{\kappa_{b}}{2}(t-t^{\prime})}(-iG_{2}\hat{c}(t^{\prime})-iG_{nv}\hat{d}(t^{\prime}))dt^{\prime}  \nonumber
\end{cases}
\end{eqnarray}
The task is to evaluate the integral $X=\int_{-\infty}^{t}e^{\frac{-\kappa (t-t^{\prime})}{2}}x(t^{\prime})dt^{\prime}$. Note that the function:
\begin{eqnarray}
\mu_{\kappa}=\frac{\kappa}{4}e^{-\kappa |t|/2} \nonumber
\end{eqnarray}
will converge to the Dirac function in the limit of $\kappa \to \infty$. With the product of a Dirac distribution and Heaviside function, we arrive at the relation:
\begin{eqnarray}
\lim_{\kappa \to \infty}X=\lim_{\kappa \to \infty}\int_{-\infty}^{t}e^{\frac{-\kappa (t-t^{\prime})}{2}}x(t^{\prime})dt^{\prime}  \approx \frac{2}{\kappa}x(t) \nonumber
\end{eqnarray}
Then the dynamics of operator $\hat{a}$ and $\hat{b}$ will have a simpler form:
\begin{eqnarray}
\begin{cases}
\hat{a}(t) = \frac{2}{\kappa_{a}}(-iG_{1}\hat{c}(t)) \\
\hat{b}(t) = \frac{2}{\kappa_{b}}(-iG_{2}\hat{c}(t)-iG_{nv}\hat{d}(t)) \nonumber
\end{cases}
\end{eqnarray}
Plugging these expressions into the the last two equations in Eq.~(\ref{HeisenbergEqu}), we reach  Eq.~(\ref{radiation damping}) in the main text. 
\begin{eqnarray}
\begin{cases}
\partial_{t}\hat{c}  = -\frac{2G_{2}G_{nv}}{\kappa_{b}}\hat{d} - (\frac{2G_{1}^{2}}{\kappa_{a}}+\frac{2G_{2}^{2}}{\kappa_{b}}+\frac{\gamma_{c}}{2})\hat{c} \equiv -A_{cc}\hat{c}-A_{cd}\hat{d} \\
\partial_{t}\hat{d} =  -\frac{2G_{2}G_{nv}}{\kappa_{b}}\hat{c} - (\frac{2G_{nv}^{2}}{\kappa_{b}}+\frac{\gamma_{d}}{2}) \hat{d}\equiv -A_{dc}\hat{c}-A_{dd}\hat{d} \nonumber
\end{cases}
\end{eqnarray}
One can then simply get the dynamics of the two operators:
\begin{eqnarray}
\begin{cases}
\hat{c}(t) = \hat{c}(0)e^{(-A_{cd}\nu+A_{cc})} \\
\hat{b}(t) = \hat{d}(0)e^{(-A_{dc}/\nu+A_{dd})} \nonumber
\end{cases}
\end{eqnarray}
with $\nu=[(A_{dd}-A_{cc})\pm ((A_{cc}-A_{dd})+4A_{cd}^{2})^{1/2}]/2A_{cd}$. In the case of $A_{cc}=A_{dd}$, the two spin ensemble modes will decay at the same rate.

\section{Spin ensemble interactions mediated by virtual photons}\label{virtual}
Assume the information is already stored in the REDC spin ensemble and here we consider its interaction with the NV spins, mediated by the microwave resonator that is far away detuned, with detuning $\Delta_{mw}$. In the one-excitation basis of each mode, the effective Hamiltonian of the REDC-microwave-NV subsystem can be written as:
\begin{eqnarray}
H_{sub}= \begin{pmatrix}
   0 & G_{2} & 0 \\
   G_{2} & \Delta_{mw}  & G_{nv} \\
   0 & G_{nv} & 0
  \end{pmatrix}\nonumber
\end{eqnarray}
In the limit of $\Delta_{mw}\gg G_{2}, G_{nv} $, the microwave photons are virtually populated and we may take the coupling terms as perturbations, and reach the approximate Hamiltonian in the one-excitation basis of the two ensembles:
\begin{eqnarray}
H_{sub}= \frac{1}{\Delta_{mw}}\begin{pmatrix}
   G_{2}^{2} & G_{nv}G_{2}  \\
    G_{nv}G_{2} & G_{nv}^{2}
  \end{pmatrix}\nonumber
\end{eqnarray}
The corresponding eigenenergies and eigenstates are:
\begin{eqnarray}
E_{D}&=&0,\qquad\qquad  |D\rangle = \frac{1}{G_{tot}}(G_{2}|0,1\rangle - G_{nv}|1,0\rangle) \nonumber \\ 
E_{B}&=&-\frac{G_{tot}^{2}}{\Delta_{mw}},\quad\ |B\rangle = \frac{1}{G_{tot}}(G_{nv}|0,1\rangle + G_{2}|1,0\rangle),  \nonumber
\end{eqnarray}
with $G_{tot}=\sqrt{G_{2}^{2}+G_{nv}^{2}}$. In the large detuning case, the coupling between two spin ensembles will be sufficiently lower compared to $|G_{1}|$.
Note that if one takes the photon excitation into consideration by solving the 3-by-3 matrix,  the dark state $|D\rangle$ actually corresponds to the case of zero microwave excitation, while the bright state $|B\rangle$ has a term with one excitation, but it is suppressed by a factor $\frac{G_{tot}}{\Delta_{mw}}$. 


\section{Entanglement generation}\label{LossyEntanglementProtocol}
Here we show how entanglement could be generated between tele-photon and NV ensemble with sequential gates. Working in the one-excitation basis, at $t_{0}=0$ only the NV ensemble is in the one-excitation state and all other systems are in the vacuum state. Consider the system evolution for a time $G_{nv}t=\pi/4$ under the Hamiltonian stated in Eq. (\ref{Hnv}), while the REDC and microwave resonator are off-resonance. This corresponds to a $\sqrt{iSWAP}$ gate, which will create entanglement between microwave photon with the NV center ensemble:
\begin{eqnarray}
\begin{split}
& |\psi_{1}\rangle=U_{b,d}(t)|\psi_{0}\rangle=e^{-i(\hat{b}^{\dagger}\hat{d}+H.c)G_{nv}\pi/4}|\psi_{0}\rangle \\ 
& =  \begin{pmatrix}
   1 & 0 & 0 & 0 \\
   0 & 1/\sqrt{2}  & i/\sqrt{2} & 0 \\
   0 & i/\sqrt{2} &  1/\sqrt{2}& 0 \\
   0 & 0 & 0 & 1
    \end{pmatrix}
    \begin{pmatrix} 0 \\ 1 \\ 0 \\ 0 \end{pmatrix} =  \begin{pmatrix} 0 \\ 1/\sqrt{2}  \\ i/\sqrt{2}  \\ 0 \end{pmatrix}
      \nonumber
\end{split}
\end{eqnarray}

Then, by tuning $G_{2}\gg |G_{1}|$ and microwave resonator off resonance with the NV ensemble, the second iSWAP gate (corresponding to an interaction time $G_{2}t=\pi/2$) would transfer the state of microwave resonator to REDC, while leaving the former in the ground state. In the subspace of these two subsystems, it can be shown:
\begin{eqnarray}
\begin{split}
& |\psi_{2}\rangle=U_{c,b}(t)|\psi_{1}\rangle=e^{-i(\hat{c}^{\dagger}\hat{b}+H.c)G_{2}\pi/2}|\psi_{1}\rangle \\
& =\begin{pmatrix}
   1 & 0 & 0 & 0 \\
   0 & 0  & i & 0 \\
   0 & i &  0& 0 \\
   0 & 0 & 0 & 1
    \end{pmatrix}
    \begin{pmatrix} 1\sqrt{2} \\ i/\sqrt{2} \\ 0 \\0 \end{pmatrix} =  \begin{pmatrix}  1/\sqrt{2} \\0 \\ -1/\sqrt{2}  \\ 0 \end{pmatrix}
      \nonumber
\end{split}
\end{eqnarray}
Another iSWAP gate in succession will finally create entanglement between tele-photon and NV ensemble while leaving the REDC in its ground state:
\begin{eqnarray}
\begin{split}
& |\psi_{3}\rangle=U_{a,c}(t)|\psi_{2}\rangle=e^{-i(\hat{a}^{\dagger}\hat{c}+H.c)G_{1}\pi/2}|\psi_{2}\rangle \\
& =\begin{pmatrix}
   1 & 0 & 0 & 0 \\
   0 & 0  & i & 0 \\
   0 & i &  0& 0 \\
   0 & 0 & 0 & 1
    \end{pmatrix}
    \begin{pmatrix} 1\sqrt{2} \\ -1/\sqrt{2} \\ 0 \\0 \end{pmatrix} =  \begin{pmatrix}  1/\sqrt{2} \\0 \\ -i/\sqrt{2}  \\ 0 \end{pmatrix}
      \nonumber
\end{split}
\end{eqnarray}

This procedure could be reversed by starting with a number state in optical cavity. However, as shown in the figure below, the cavity decay would only yield a final NV-optical entanglement concurrence of 0.28, which means it is unlikely to success under the aforementioned decay parameters. 

\begin{figure}[h]
\centering
\includegraphics[scale=0.45]{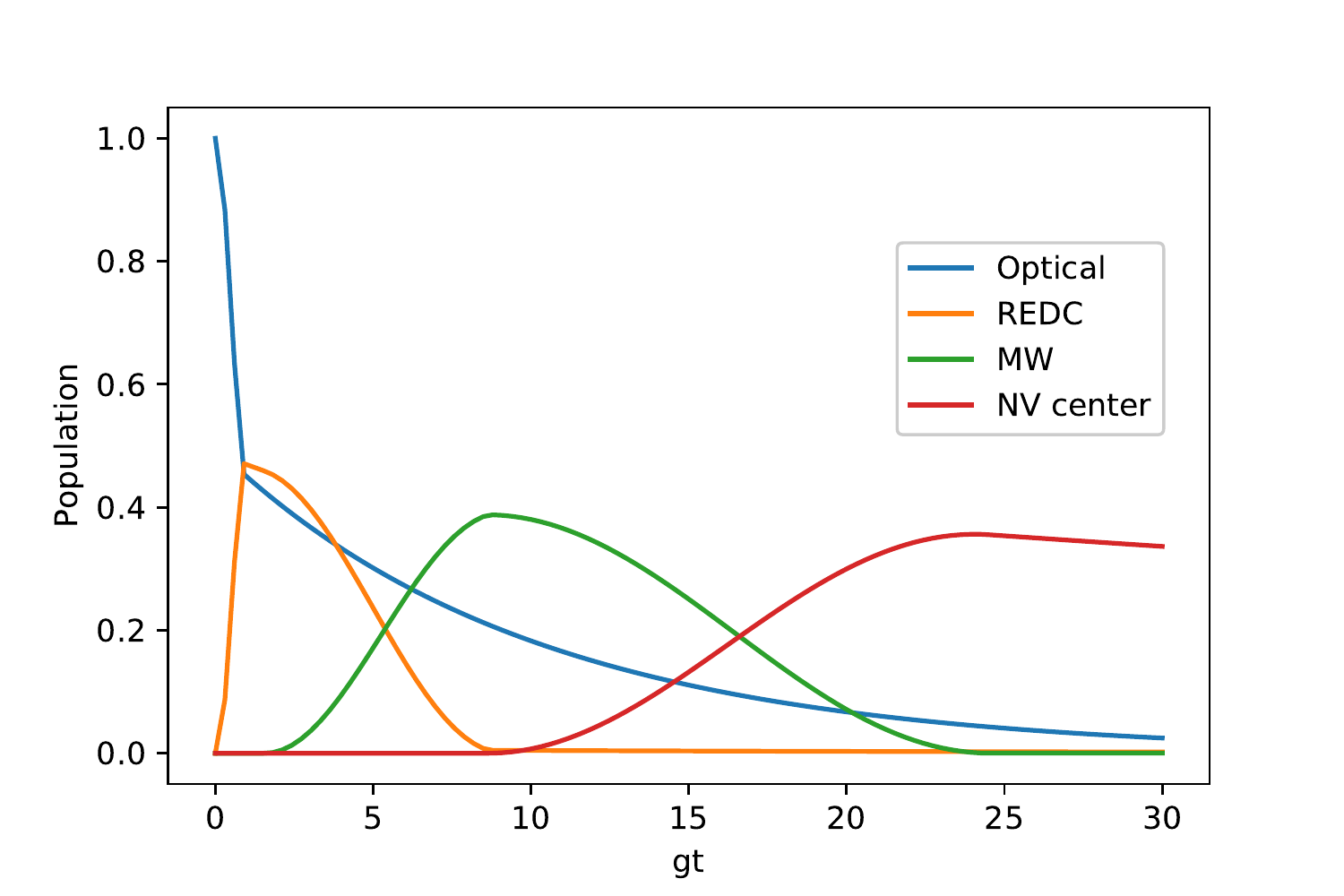}
\includegraphics[scale=0.45]{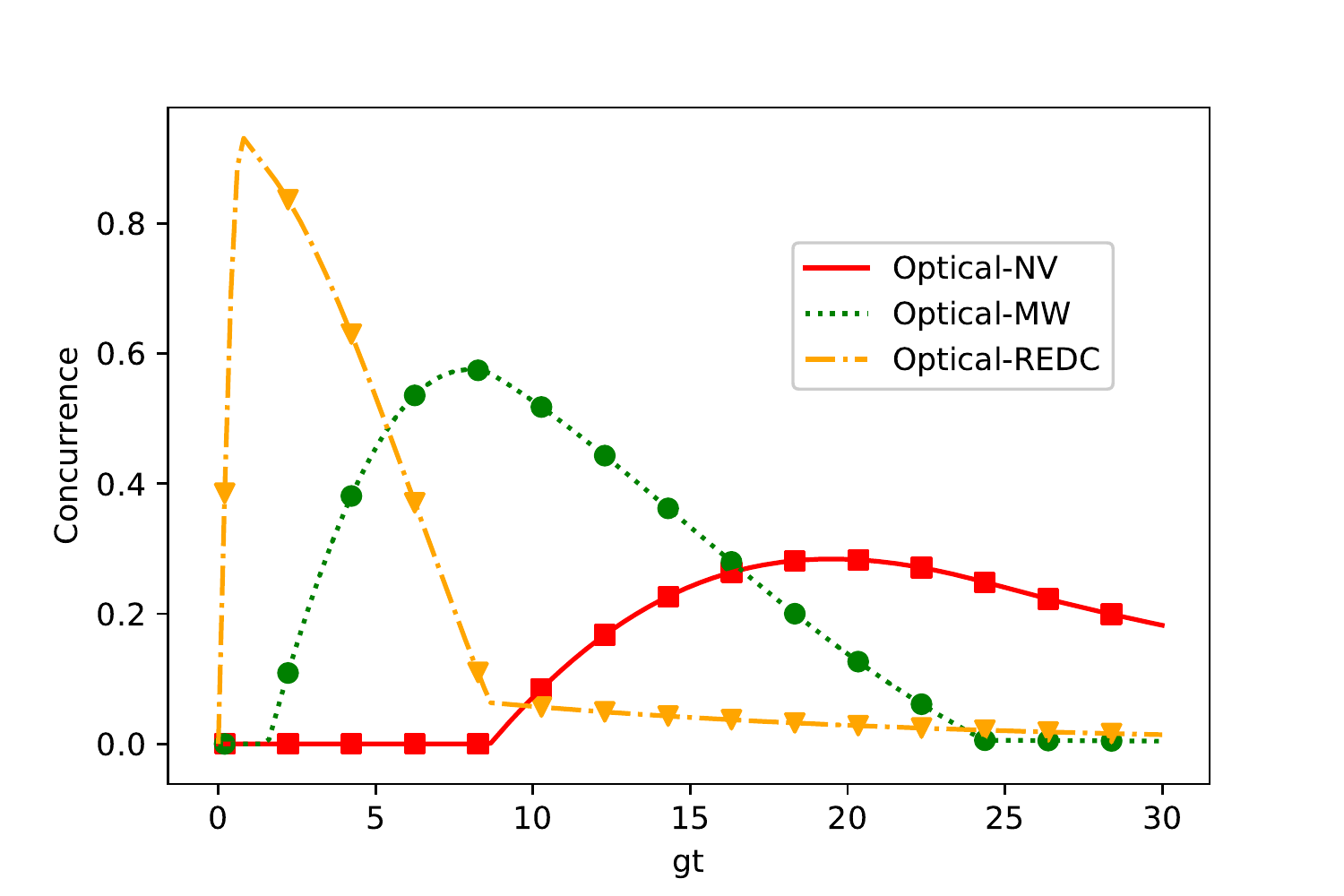}
\caption{Entanglement generation between tele-photon and NV spin ensemble assuming photon state is initially populated. \textit{Top:} Subsystem population as a function of time.  \textit{Bottom:} Concurrence between optical photon and other subsystems as a function of time. NV centers and optical photons  can only have a concurrence of 0.28. Again, the coupling strengths are $|G_{1}|=5|G_{2}|=10G_{nv}=g$ and the decay parameters are $\kappa_{a}=2.5\gamma_{c}=10\gamma_{nv}=100\kappa_{b}=0.1g$.   }
\end{figure}


\nocite{*}

\bibliography{NVtele_arXiv}

\end{document}